\documentclass[12pt]{article}
\usepackage{epsfig}
\usepackage{amsmath}
\usepackage{hhline}
\usepackage{amssymb}
\usepackage{times}
\usepackage{cite}

\newlength{\dinwidth}
\newlength{\dinmargin}
 \usepackage{ifpdf}

 \usepackage{thumbpdf}
\usepackage[
        colorlinks=false,
        urlcolor=rltblue,       
        filecolor=rltgreen,     
        linkcolor=rltred,       
        pdftitle={Measurement of Inclusive Jet Production in Deep-Inelastic Scattering at High Q2 and Determination of the Strong Coupling},
        pdfauthor={H1 Collaboration},
        pdfsubject={Inclusive Jet Cross Section},
        pdfkeywords={High-Energy Physics, Particle Physics},
        pdfpagemode=None,
        bookmarksopen=true]{hyperref}

\usepackage{color}
\definecolor{rltred}{rgb}{0.75,0,0}
\definecolor{rltgreen}{rgb}{0,0.5,0}
\definecolor{rltblue}{rgb}{0,0,0.75}
\begin{document}  

\newcommand{\pb}{\,\rm pb}
\newcommand{\cm}{\,\rm cm}
\newcommand{\pom}{{I\!\!P}}
\newcommand{\reg}{{I\!\!R}}
\newcommand{\slowpi}{\pi_{\mathit{slow}}}
\newcommand{\fiidiii}{F_2^{D(3)}}
\newcommand{\fiidiiiarg}{\fiidiii\,(\beta,\,Q^2,\,x)}
\newcommand{\n}{1.19\pm 0.06 (stat.) \pm0.07 (syst.)}
\newcommand{\nz}{1.30\pm 0.08 (stat.)^{+0.08}_{-0.14} (syst.)}
\newcommand{\fiidiiiful}{F_2^{D(4)}\,(\beta,\,Q^2,\,x,\,t)}
\newcommand{\fiipom}{\tilde F_2^D}
\newcommand{\ALPHA}{1.10\pm0.03 (stat.) \pm0.04 (syst.)}
\newcommand{\ALPHAZ}{1.15\pm0.04 (stat.)^{+0.04}_{-0.07} (syst.)}
\newcommand{\fiipomarg}{\fiipom\,(\beta,\,Q^2)}
\newcommand{\pomflux}{f_{\pom / p}}
\newcommand{\nxpom}{1.19\pm 0.06 (stat.) \pm0.07 (syst.)}
\newcommand {\gapprox}
   {\raisebox{-0.7ex}{$\stackrel {\textstyle>}{\sim}$}}
\newcommand {\lapprox}
   {\raisebox{-0.7ex}{$\stackrel {\textstyle<}{\sim}$}}
\def\gsim{\,\lower.25ex\hbox{$\scriptstyle\sim$}\kern-1.30ex%
\raise 0.55ex\hbox{$\scriptstyle >$}\,}
\def\lsim{\,\lower.25ex\hbox{$\scriptstyle\sim$}\kern-1.30ex%
\raise 0.55ex\hbox{$\scriptstyle <$}\,}
\newcommand{\pomfluxarg}{f_{\pom / p}\,(x_\pom)}
\newcommand{\dsf}{\mbox{$F_2^{D(3)}$}}
\newcommand{\dsfva}{\mbox{$F_2^{D(3)}(\beta,Q^2,x_{I\!\!P})$}}
\newcommand{\dsfvb}{\mbox{$F_2^{D(3)}(\beta,Q^2,x)$}}
\newcommand{\dsfpom}{$F_2^{I\!\!P}$}
\newcommand{\gap}{\stackrel{>}{\sim}}
\newcommand{\lap}{\stackrel{<}{\sim}}
\newcommand{\fem}{$F_2^{em}$}
\newcommand{\tsnmp}{$\tilde{\sigma}_{NC}(e^{\mp})$}
\newcommand{\tsnm}{$\tilde{\sigma}_{NC}(e^-)$}
\newcommand{\tsnp}{$\tilde{\sigma}_{NC}(e^+)$}
\newcommand{\st}{$\star$}
\newcommand{\sst}{$\star \star$}
\newcommand{\ssst}{$\star \star \star$}
\newcommand{\sssst}{$\star \star \star \star$}
\newcommand{\tw}{\theta_W}
\newcommand{\sw}{\sin{\theta_W}}
\newcommand{\cw}{\cos{\theta_W}}
\newcommand{\sww}{\sin^2{\theta_W}}
\newcommand{\cww}{\cos^2{\theta_W}}
\newcommand{\trm}{m_{\perp}}
\newcommand{\trp}{p_{\perp}}
\newcommand{\trmm}{m_{\perp}^2}
\newcommand{\trpp}{p_{\perp}^2}
\newcommand{\alp}{\alpha_s}

\newcommand{\alps}{\alpha_s}
\newcommand{\alpset}{\alpha_s(E_T)}
\newcommand{\sqrts}{$\sqrt{s}$}
\newcommand{\LO}{$O(\alpha_s^0)$}
\newcommand{\Oa}{$O(\alpha_s)$}
\newcommand{\Oaa}{$O(\alpha_s^2)$}
\newcommand{\PT}{p_{\perp}}
\newcommand{\JPSI}{J/\psi}
\newcommand{\sh}{\hat{s}}
\newcommand{\uh}{\hat{u}}
\newcommand{\MP}{m_{J/\psi}}
\newcommand{\PO}{I\!\!P}
\newcommand{\xbj}{x}
\newcommand{\xpom}{x_{\PO}}
\newcommand{\ttbs}{\char'134}
\newcommand{\xpomlo}{3\times10^{-4}}  
\newcommand{\xpomup}{0.05}  
\newcommand{\dgr}{^\circ}
\newcommand{\pbarnt}{\,\mbox{{\rm pb$^{-1}$}}}
\newcommand{\gev}{\,\mbox{GeV}}
\newcommand{\WBoson}{\mbox{$W$}}
\newcommand{\fbarn}{\,\mbox{{\rm fb}}}
\newcommand{\fbarnt}{\,\mbox{{\rm fb$^{-1}$}}}
\newcommand{\dsdx}[1]{$d\sigma\!/\!d #1\,$}
\newcommand{\eV}{\mbox{e\hspace{-0.08em}V}}
%
%
\newcommand{\qsq}{\ensuremath{Q^2} }
\newcommand{\gevsq}{\ensuremath{\mathrm{GeV}^2} }
\newcommand{\et}{\ensuremath{E_t^*} }
\newcommand{\rap}{\ensuremath{\eta^*} }
\newcommand{\gp}{\ensuremath{\gamma^*}p }
\newcommand{\dsiget}{\ensuremath{{\rm d}\sigma_{ep}/{\rm d}E_t^*} }
\newcommand{\dsigrap}{\ensuremath{{\rm d}\sigma_{ep}/{\rm d}\eta^*} }

\newcommand{\dstar}{\ensuremath{D^*}}
\newcommand{\dstarp}{\ensuremath{D^{*+}}}
\newcommand{\dstarm}{\ensuremath{D^{*-}}}
\newcommand{\dstarpm}{\ensuremath{D^{*\pm}}}
\newcommand{\zDs}{\ensuremath{z(\dstar )}}
\newcommand{\Wgp}{\ensuremath{W_{\gamma p}}}
\newcommand{\ptds}{\ensuremath{p_t(\dstar )}}
\newcommand{\etads}{\ensuremath{\eta(\dstar )}}
\newcommand{\ptj}{\ensuremath{p_t(\mbox{jet})}}
\newcommand{\ptjn}[1]{\ensuremath{p_t(\mbox{jet$_{#1}$})}}
\newcommand{\etaj}{\ensuremath{\eta(\mbox{jet})}}
\newcommand{\detadsj}{\ensuremath{\eta(\dstar )\, \mbox{-}\, \etaj}}

\def\Journal#1#2#3#4{{#1} {\bf #2} (#3) #4}
\def\NCA{\em Nuovo Cimento}
\def\NIM{\em Nucl. Instrum. Methods}
\def\NIMA{{\em Nucl. Instrum. Methods} {\bf A}}
\def\NPB{{\em Nucl. Phys.}   {\bf B}}
\def\PLB{{\em Phys. Lett.}   {\bf B}}
\def\PRL{\em Phys. Rev. Lett.}
\def\PRD{{\em Phys. Rev.}    {\bf D}}
\def\ZPC{{\em Z. Phys.}      {\bf C}}
\def\EJC{{\em Eur. Phys. J.} {\bf C}}
\def\CPC{\em Comp. Phys. Commun.}

\hyphenation{QCD-Compton}

\begin{titlepage}
\begin{flushleft}
{\tt DESY 07-073    \hfill    ISSN 0418-9833} \\
{\tt May  2007}                  \\
\end{flushleft}
\vspace{2cm}
\begin{center}
\begin{Large}

{\bf \boldmath Measurement of Inclusive Jet Production in\\ Deep-Inelastic Scattering at High $Q^2$ and Determination of the Strong Coupling }

\vspace{2cm}

H1 Collaboration

\end{Large}
\end{center}

\vspace{2cm}

\begin{abstract}
\noindent
Inclusive jet production is studied in neutral
current deep-inelastic positron-proton scattering at large
four momentum transfer squared $Q^2>150\gev^2$ with the H1
detector at HERA.
Single and double differential inclusive jet cross sections are measured
 as a function of $Q^2$ and of the transverse energy
$E_T$ of the jets in the Breit frame. 
The measurements are found to be well described by
calculations at next-to-leading order in perturbative QCD.
 The running of the strong coupling is demonstrated  
and the value of $\alpha_s(M_Z)$ is determined.
The ratio of the inclusive jet cross section to the inclusive neutral
current cross section is also measured and used to extract a precise
value for $\alpha_s(M_Z) = 0.1193 ~\pm 0.0014\,\mathrm{(exp.)}
~ ^{+0.0047}_{-0.0030}\,\mathrm{(th.)}~ \pm 0.0016\,\mathrm{(pdf)}~ $.
\end{abstract}

\vspace{1.5cm}

\begin{center}
Submitted to \PLB
\end{center}

\end{titlepage}

\begin{flushleft}

A.~Aktas$^{11}$,               
C.~Alexa$^{5}$,                
V.~Andreev$^{25}$,             
T.~Anthonis$^{4}$,             
B.~Antunovic$^{26}$,           
S.~Aplin$^{11}$,               
A.~Asmone$^{33}$,              
A.~Astvatsatourov$^{4}$,       
S.~Backovic$^{30}$,            
A.~Baghdasaryan$^{38}$
% ...

\bigskip{\it
% ...
$ ^{4}$ Inter-University Institute for High Energies ULB-VUB, Brussels;
          Universiteit Antwerpen, Antwerpen; Belgium$^{ c}$ \\
 $ ^{5}$ National Institute for Physics and Nuclear Engineering (NIPNE) ,
          Bucharest, Romania \\
% ...
$ ^{11}$ DESY, Hamburg, Germany \\
% ...
$ ^{25}$ Lebedev Physical Institute, Moscow, Russia$^{ e}$ \\
 $ ^{26}$ Max-Planck-Institut f\"ur Physik, M\"unchen, Germany \\
% ...
$ ^{30}$ Faculty of Science, University of Montenegro,
          Podgorica, Montenegro$^{ e}$ \\
% ...
$ ^{33}$ Dipartimento di Fisica Universit\`a di Roma Tre
          and INFN Roma~3, Roma, Italy \\
% ...
$ ^{38}$ Yerevan Physics Institute, Yerevan, Armenia \\
% ...
$ ^c$ Supported by FNRS-FWO-Vlaanderen, IISN-IIKW and IWT
      and  by Interuniversity
Attraction Poles Programme,
      Belgian Science Policy \\
% ...
$ ^e$ Supported by the Deutsche Forschungsgemeinschaft \\
% ...
}
\end{flushleft}

\newpage


\section{Introduction}

\begin{figure}[b]
\centering
 \includegraphics[height=4.4cm]{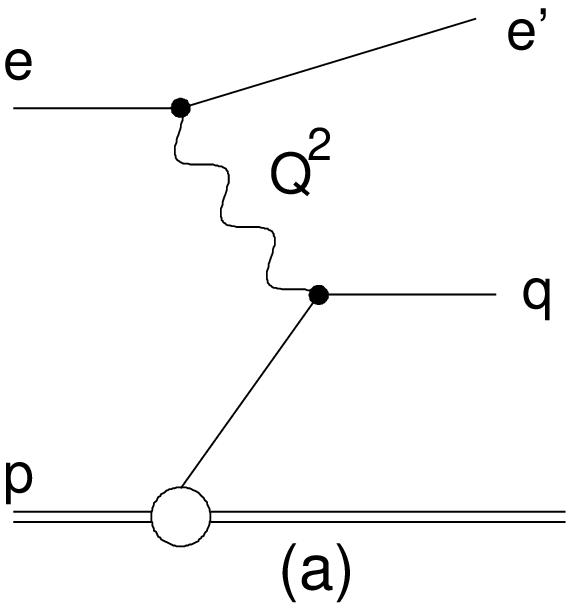}\hskip1.0cm
 \includegraphics[height=4.4cm]{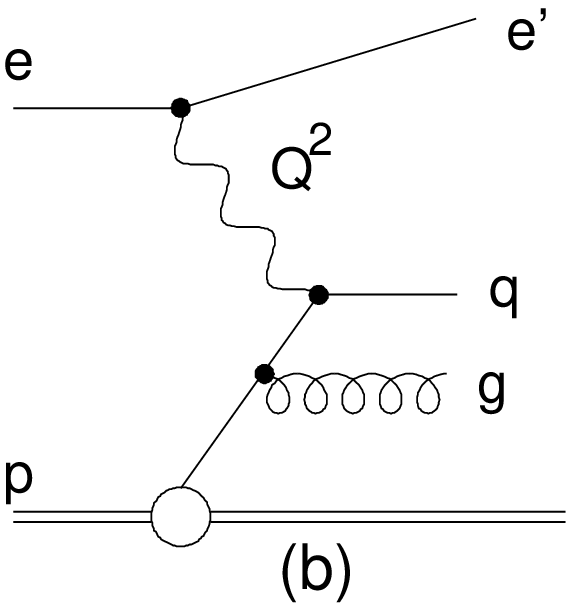}\hskip1.0cm
 \includegraphics[height=4.4cm]{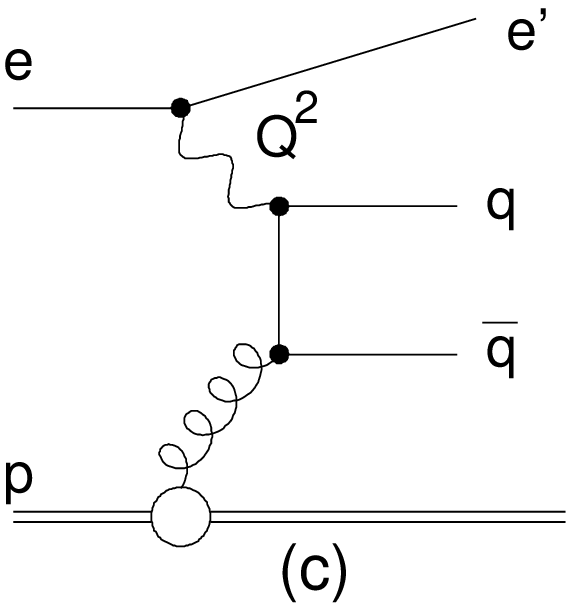}
\caption[Diagrams of different order in $\alpha_s$ in deep-inelastic
lepton-proton scattering]%
{
Deep-inelastic lepton-proton scattering at different orders in
$\alpha_s$: (a) Born contribution $\mathcal{O}(1)$, (b) QCD Compton
    scattering $\mathcal{O}(\alpha_s)$ and (c) boson-gluon fusion $\mathcal{O}(\alpha_s)$.
}
\label{fig:feynborn}
\end{figure}

Jet production in neutral current (NC) deep-inelastic
scattering (DIS) at HERA provides an important testing ground for Quantum Chromodynamics (QCD).
The Born contribution in DIS (figure~\ref{fig:feynborn}a)
gives only indirect information on the strong coupling $\alpha_s$ 
via scaling violations of the proton structure functions.
At leading order (LO) in $\alpha_s$ additional processes contribute:
QCD-Compton (figure~\ref{fig:feynborn}b) 
and boson-gluon fusion (figure~\ref{fig:feynborn}c).
In the Breit frame of reference~\cite{feynman,Adloff:2000tq},
where the virtual boson and the proton collide head on,
 the Born contribution generates no transverse momenta.
Partons with transverse momenta are produced in lowest order by
 the QCD-Compton and boson-gluon fusion processes.
Jet production in the Breit frame therefore provides direct sensitivity to
 $\alpha_s$ and allows for a precision test of QCD.

Analyses of inclusive jet production in DIS at high four momentum transfer squared $Q^2$
were previously performed by the H1\cite{Adloff:2000tq} and ZEUS\cite{Chekanov:2002be,Chekanov:2006xr} collaborations at HERA.
Perturbative QCD (pQCD) calculations supplemented with hadronisation corrections were found to describe the data.
The strong coupling $\alpha_s$ and the gluon density in the proton were both extracted.

In this paper new measurements of the inclusive jet cross section are presented, 
 based on data corresponding to  twice the integrated luminosity and
a higher centre-of-mass energy than in the previous H1 analysis\cite{Adloff:2000tq}.
The larger data set together with improved understanding of the hadronic energy measurement significantly reduces the total 
uncertainty of the results.
Differential inclusive jet cross sections are measured as functions of the hard scales $Q^2$ and the transverse jet energy $E_T$ in the Breit frame
in the ranges $150<Q^2<15000\gev^2$ and $7<E_T<50\gev$.
In addition, the ratio of the jet cross section to the inclusive NC DIS cross section,
in the following referred to as the normalised inclusive jet cross section, is determined.
This observable benefits from a partial cancellation of  experimental and theoretical uncertainties.
The measurements are compared with pQCD predictions at next-to-leading order (NLO),
 and the strong coupling $\alpha_s$ is determined from a fit of the predictions to the data.

\section{Experimental Method}
\label{sec:expmethod}
The data were collected with the H1 detector at HERA in the years 1999 and 2000.
During this period HERA collided positrons of energy $E_e=27.5\gev$ with protons of energy \mbox{$E_p=920\gev$}
giving a centre-of-mass energy $\sqrt{s} = 319\gev$.
The data sample used in this analysis corresponds to an integrated luminosity of $65.4\pb^{-1}$.

\subsection{H1 detector}
A detailed description of the H1 detector can be found 
in~\cite{Abt:1996hi,Appuhn:1996na}.
H1 uses a right-handed coordinate system with the origin at the nominal interaction point and 
the $z$-axis along the beam direction, the $+z$ or ``forward" direction being that of the outgoing proton beam.
Polar angles $\theta$ and azimuthal angles $\phi$ are defined with respect to this axis.
The pseudorapidity is related to the polar angle  $\theta$ by $\eta=-\mathrm{ln}\,\mathrm{tan}(\theta/2)$.
The detector components important for this analysis are described below.

The electromagnetic and hadronic energies are measured
using the Liquid Argon (LAr) ca\-lo\-ri\-me\-ter
in the polar angular range $4^\circ <\theta <154^\circ$
and with full azimuthal coverage.
The LAr calorimeter consists of an electromagnetic section 
($20$~to~$30$ radiation lengths) with lead absorbers and a 
hadronic section with steel absorbers.
The total depth of both sections varies between $4.5$ and $8$ 
interaction lengths.
The energy resolution is $\sigma_E/E = 12\%/\sqrt{E \;/\gev}\oplus 1\%$
for electrons and 
$\sigma_{E}/E = 50\%/\sqrt{E\;/\gev}\oplus 2\%$
for hadrons, as obtained from test beam measurements~\cite{Andrieu:1993tz}.
In the backward region ($153\dgr\le\theta\le 177\dgr$) energy is measured
by a lead/scintillating fibre Spaghetti-type Calorimeter (SpaCal) composed
of an electromagnetic and a hadronic section.
The energy resolution of the SpaCal is $\sigma_E/E \approx 7\%/\sqrt{E \;/\gev}\oplus 1\%$
 for electrons~\cite{Nicholls:1995di}.
The central tracking system ($20\dgr\le\theta\le 160\dgr$) is located inside
 the LAr calorimeter  and consists of drift and proportional chambers,
complemented by a silicon vertex detector~\cite{Pitzl:2000wz} covering the range $30\dgr\le\theta\le 150\dgr$. 
The chambers and calorimeters are surrounded by a superconducting
solenoid providing a uniform field of $1.16\,\mathrm{T}$ inside the tracking
volume. 

The scattered positron is identified as an electromagnetic cluster in the LAr calorimeter 
with an associated track.
The remaining clusters in the calorimeters and charged tracks are attributed to the 
hadronic final state which is reconstructed using an energy flow algorithm that 
avoids double counting of energy.                                                                                                                          
The luminosity is determined by measuring the Bethe-Heitler 
process (\mbox{$ep\rightarrow ep\gamma$}), where the photon is detected in a 
calorimeter close to the beam pipe at $z=-103\,\mathrm{m}$.

\subsection{Event and jet selection}
NC DIS events are selected by requiring the scattered positron to be
detected in the LAr calorimeter with a reconstructed energy $E'_{e}$ exceeding  
$11~\gev$ and a polar angle $\theta_\mathrm{e} < 153^o$.
These requirements ensure a trigger efficiency of greater than $98\%$.
The \mbox{$z$-coordinate} of the event vertex is required to be within $\pm 35\cm$ 
of the average position of the interaction point. This condition reduces contributions 
from beam induced background and cosmic muons.
Non-$ep$ background is further reduced by requiring an event timing which 
matches the HERA bunch crossing.
The total longitudinal energy balance must satisfy $45 < \sum_i
(E_i-p_{z,i}) < 65~\gev$, where the sum runs over all detected particles.
This requirement reduces the contributions of the photoproduction background and
of DIS with initial state photon radiation
for which the escaped positron or photon in the $-z$-direction leads to values of this
observable lower than the expectation $2E_e=55\gev$, for events with losses only along the 
outgoing proton beam.
Elastic QED Compton and lepton pair production processes are suppressed by rejecting events
containing a second isolated electromagnetic deposit and no hadronic activity.
The remaining photoproduction background is estimated
using Monte Carlo simulations and found to be negligible in all $Q^2$ and jet $E_T$ bins.

The DIS phase space covered by this analysis is defined by
$$ 150<Q^2<15000~\gev^2 \ ,$$
$$0.2<y<0.7 \ ,$$
where $y$,   ,quantifies the inelasticity of the interaction.
These two variables are reconstructed from the four momenta of the scattered positron and the
hadronic final state particles using the electron-sigma method~\cite{Bassler:1994uq}.

The jet analysis is performed in the Breit frame.
The boost from the laboratory system to the Breit frame is determined by $Q^2$, $y$ and
 the azimuthal angle of the scattered positron.
Particles of the hadronic final state are clustered
into jets using the inclusive $k_T$ algorithm~\cite{Ellis:1993tq}
with the $p_T$ recombination scheme and with distance parameter $R=1$ in the \mbox{$\eta$-$\phi$} plane.
The cut  $-1.0 < \eta^\mathrm{Lab} < 2.5$ ensures that jets are
 well contained within the acceptance of the LAr calorimeter, where $\eta^\mathrm{Lab}$
is the jet pseudorapidity in the laboratory frame.
Every jet with $7 < E_T < 50\gev$ contributes to the inclusive jet cross section,
 regardless of the jet multiplicity in the event.
In total 23714 jets pass the analysis cuts.

In addition, the normalised inclusive jet cross section is investigated,
calculated as the ratio of the number of jets to the number of selected NC DIS events
in the $y$~range defined above.
This observable equals the average jet multiplicity of NC DIS events within the given phase space.
Jet cross sections and normalised jet cross sections are studied as a function of $Q^2$ and $E_T$.

\subsection{Cross section determination}
\label{sec:datacorrrection}
In order to extract the cross sections at hadron level, the experimental data are corrected for limited detector acceptance
and resolution.
The correction factors are determined using simulated NC~DIS events.
The generated events are passed through a detailed simulation of the H1 detector and subjected to the
same reconstruction and analysis chain as the data.
The following Monte Carlo event generators are used for this purpose: DJANGOH\cite{Charchula:1994kf} using the
 Color Dipole Model as implemented in  ARIADNE\cite{Lonnblad:1992tz}, and RAPGAP\cite{Jung:1993gf}
 using matrix elements matched with parton showers.
Both RAPGAP and DJANGOH provide a good description of the data in both the inclusive and the jet sample.
The purity of the jet sample, defined as the fraction of events reconstructed in a bin that originate from that bin on
 hadron level, is found to be larger than $60\%$ in all analysis bins.
Correction factors are determined as the ratio of the cross section obtained from particles at hadron level
to the cross section calculated using particles reconstructed in the detector.
This correction is applied bin-by-bin in $Q^2$ and $E_T$.
Arithmetic means of the correction factors determined by RAPGAP and DJANGOH are used, and
 half of the difference is assigned as model uncertainty.
The correction factors deviate typically by less than $20\%$ from unity.
The effects of QED radiation are corrected for using the HERACLES~\cite{Kwiatkowski:1990es} program.
The size of these corrections is typically $10\%$ for the jet cross sections and $~5\%$
 for the normalised jet cross sections.

For the normalised jet cross sections the ratio of the number of jets to the number of NC DIS events is calculated on detector level,
and it is this ratio which is corrected for detector and QED effects.

\subsection{Systematic errors}
The following sources of systematic uncertainty are considered:
\begin{itemize}
\item The positron energy uncertainty is $0.7\%$~to~$3\%$ depending
  on the $z$-impact point of the positron in the calorimeter.
  Uncertainties in the positron reconstruction affect the event kinematics and
  thus the boost to the Breit frame.
  The resulting uncertainty on the cross sections and normalised cross sections is typically $0.5\%$. 
\item The positron polar angle systematic uncertainty is between $1$~and~$3~\mathrm{mrad}$.
The resulting uncertainty on the cross sections and normalised cross sections is typically $0.5\%$.
\item 
  The energy scale uncertainty of the reconstructed hadronic final state is estimated to be $2\%$,
 dominated by the uncertainty of the LAr hadronic energy scale.
  The resulting uncertainty on the cross sections and normalised cross sections is typically in the range $1$~to~$4\%$.
\item The luminosity measurement uncertainty leads to an overall 
  normalisation error of $1.5\%$ for the jet cross sections.
\item The model dependence of the data correction is estimated as described in section~\ref{sec:datacorrrection}.
 It is below $10\%$ in most of the bins and typically $2\%$.
\item An error of $1\%$ is estimated from the uncertainty of the QED radiative correction \cite{Adloff:2000qj}.
\end{itemize}

\noindent
The dominant experimental uncertainties on the jet cross section arise from the model dependence of the data correction
 and from the LAr hadronic energy scale uncertainty.
 The individual contributions are added in 
quadrature to obtain the total systematic uncertainty.
The correlations of the errors among the different bins are treated using the same procedure as
described in~\cite{Adloff:2000tq}.
The uncertainties of the luminosity measurement and of the positron polar angle are
each assumed to be fully correlated between the bins.
The error on the positron polar angle and the QED radiative corrections is assumed to be uncorrelated.
The remaining sources of systematics, namely the positron energy scale, the hadronic
final state energy scale and the model dependence are equally shared between 
correlated and uncorrelated parts.
For the normalised jet cross sections
systematic uncertainties are reduced and the luminosity uncertainty cancels.

\section{NLO QCD Calculation}
\label{sec:nlo}
Reliable quantitative predictions of jet cross sections in DIS require the perturbative
calculations to be performed to at least next-to-leading order of the strong coupling.
In order to compare with data, hadronisation corrections have to be applied to the perturbative calculations.
By using the inclusive $k_T$ jet algorithm~\cite{Ellis:1993tq} the observables
 in the present analysis are infrared and collinear safe and the hadronisation corrections are small.
In addition, by applying this algorithm in the Breit frame, jet cross sections can be calculated in pQCD,
since initial state singularities can be absorbed in the
definition of the proton parton densities.

The theoretical prediction for the jet cross section is obtained using the  NLOJET++ program~\cite{Nagy:2001xb}, which
performs the matrix element integration at NLO of the strong coupling, $\mathcal{O}(\alpha_s^2)$.
The strong coupling is taken as $\alpha_s(M_Z) = 0.118$  and is evolved as a function of the renormalisation scale at two loop precision.
The calculations are performed in the $\overline{\mbox{\rm MS}}$ scheme for five massless quark flavours.
The parton density functions (PDFs) of the proton are taken from the CTEQ6.5M set~\cite{Tung:2006tb}.
The factorisation scale $\mu_f$ is chosen to be $Q$  and the renormalisation scale $\mu_r$ is chosen to be the $E_T$ of each jet. 
Running of the electromagnetic coupling with $Q^2$ is taken into account.
No QED radiation is included in the calculation since the data are corrected for this effect.
Electroweak effects due to $Z^0$ boson exchange are determined using the LEPTO event generator \cite{Ingelman:1996mq} and
are applied as correction factors to the calculation.

The hadronisation correction factor is calculated for each bin as the ratio of the cross section defined
 at hadron level to the cross section defined at parton level.
These correction factors are determined with the same Monte Carlo event samples
 used to correct the data from detector to hadron level.
The correction factors applied to the perturbative calculations are calculated as the average of the values
from DJANGOH and RAPGAP, as described in section~\ref{sec:datacorrrection}.
The hadronisation correction factors differ typically by less than $10\%$ from unity and agree at the level
of $2\%$ between the two Monte Carlo simulations.

The theory uncertainty includes the hadronisation correction error
and the uncertainty related to the neglected higher orders in the perturbative calculation.
The systematic error attributed to the hadronisation correction is taken to be half of the difference
between the correction factors obtained using RAPGAP and DJANGOH.
The dominant uncertainty is related to the NLO accuracy and
is estimated by a variation of the
chosen scales for $\mu_r$ and $\mu_f$ by arbitrary but conventional factors
in the range from $0.5$ to $2$ applied to the nominal scales.
In seven out of the 24 bins in $Q^2$ and $E_T$ the dependence of the
pQCD calculation on $\mu_r$ is not monotone, i.e.\ the largest deviation from the central
value is found for factors within the range $0.5$ to $2$.
In such cases the difference between maximum and minimum cross sections 
 found in the variation interval is taken, in order not to underestimate the scale dependence.
Over the whole phase space, the uncertainty due to the renormalisation scale is found to be at least a factor
 of three larger than that due to the factorisation scale.
The contributions from both scale variations are added in quadrature.

In order to calculate the normalised inclusive jet cross sections,
the prediction of the inclusive jet cross section is divided by the prediction of the NC DIS cross section.
The latter is calculated at NLO, $\mathcal{O}(\alpha_s)$, with the DISENT package \cite{Catani:1996vz},
 using the same settings as for NLOJET++ and with the renormalisation and factorisation scales set to $Q$.
Again, the scale uncertainties are determined by independent variations of $\mu_r$ and $\mu_f$ 
in the range from $0.5$ to $2$ around the nominal value.
The scale uncertainties from the jet and the NC DIS part are assumed to be uncorrelated.
Consequently, the scale uncertainty for the ratio is estimated by 
adding both contributions in quadrature.
If the uncertainties are assumed to be anti correlated, which leads to the largest change,
 the resulting theory error increases only slightly by a factor of $1.15$.
The uncertainty originating from the PDFs is also taken into account.
The CTEQ6.5M set of parton densities provides variations
 which quantify the uncertainty of the central set. 
The PDF uncertainties are propagated
 into the pQCD prediction of the inclusive jet cross section and the NC DIS cross section.

The strong coupling is determined by repeating the perturbative calculations for many values of $\alpha_s(M_Z)$
 until the best match of data and theory is found.
With NLOJET++ and DISENT these calculations are time consuming.
A considerable gain in computational speed is provided by the fastNLO package~\cite{Kluge:2006xs},
which uses a two step strategy to reduce the calculation time.
In the first step, the
integration of the matrix elements is performed, which is
the time consuming part of the calculation.
This step relies for the present analysis on NLOJET++
 and DISENT and is independent of $\alpha_s(M_Z)$, PDFs and the renormalisation scale.
In the second step, the cross sections are calculated with these parameters specified.
The interpolations involved in this procedure yield a precision of better than $0.2\%$ on the cross section.
All theory calculations shown in the following are obtained using fastNLO.

\section{Results}
In the following, the differential cross sections are presented for inclusive jet production and for
normalised inclusive jet production.
Tables~\ref{tab:res_incl} and \ref{tab:res_ratio} list the measured cross sections together with their
 experimental uncertainties, separated into bin-to-bin correlated and uncorrelated parts.
These measurements are subsequently used to extract the strong coupling $\alpha_s$.

\subsection{Cross section measurements compared to NLO predictions}
\begin{figure}[ht]
\center
\includegraphics[width=1.0\textwidth]{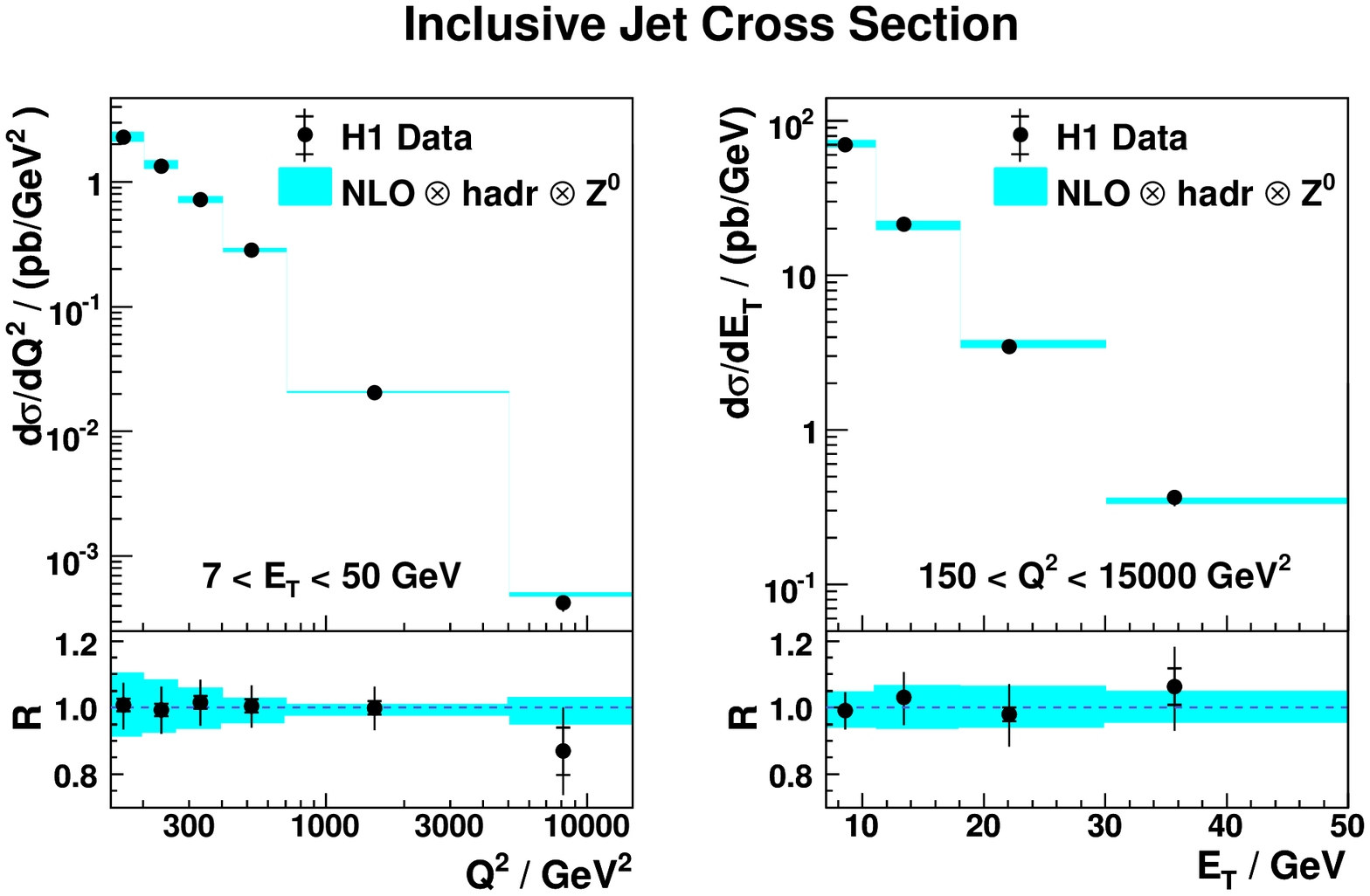}
\caption{The single differential cross section for inclusive jet production as a function of $Q^2$ (left) and of  $E_T$ (right).
The data, presented with statistical errors (inner bars)
 and total errors (outer bars), are compared with  the results of NLOJET++, corrected
for hadronisation and $Z^0$ boson exchange. 
The bands show the theoretical
uncertainty associated with the renormalisation and factorisation scales and the hadronisation correction.
In addition to the differential cross section, the ratio $R=\sigma_{\rm data}/\sigma_{\rm theory}$ is shown.
The band around $R=1$ displays the relative error of the theory calculation.
}
\label{fig:sigmasingle}
\end{figure}

\begin{figure}
\center
\includegraphics[width=1.0\textwidth]{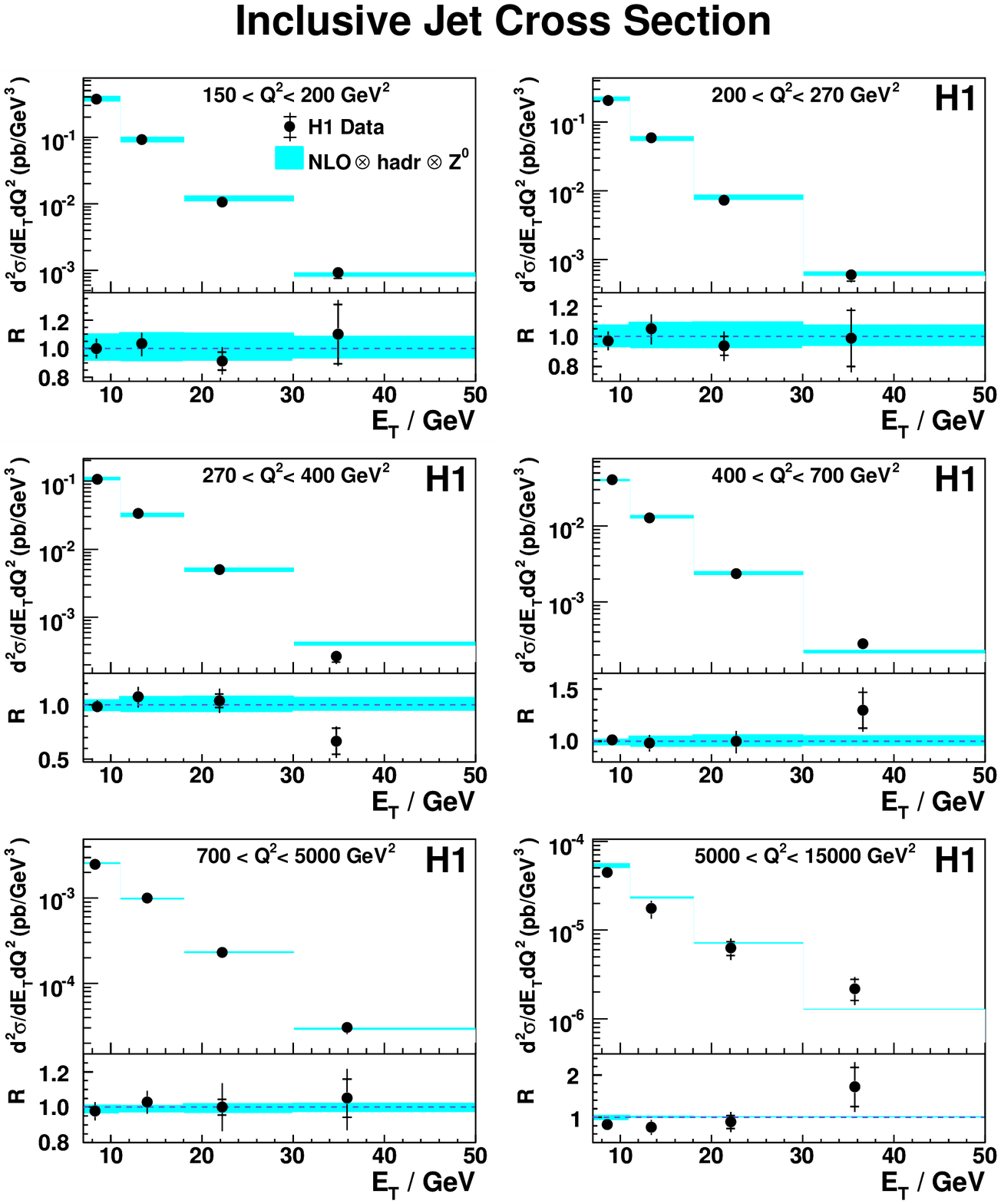}
\caption{The double differential cross section for inclusive jet production as a function of $E_T$ for six regions of $Q^2$.
The data, presented with statistical errors (inner bars)
 and total errors (outer bars), are compared with  the results of NLOJET++, corrected
for hadronisation and $Z^0$ boson exchange. 
The bands show the theoretical
uncertainty associated with the renormalisation and factorisation scales and the hadronisation correction.
In addition to the differential cross section, the ratio $R=\sigma_{\rm data}/\sigma_{\rm theory}$ is shown.
The band around $R=1$ displays the relative error of the theory calculation.
}
\label{fig:sigmadouble}
\end{figure}

\begin{figure}
\center
\vskip-1cm
\includegraphics[width=1.0\textwidth]{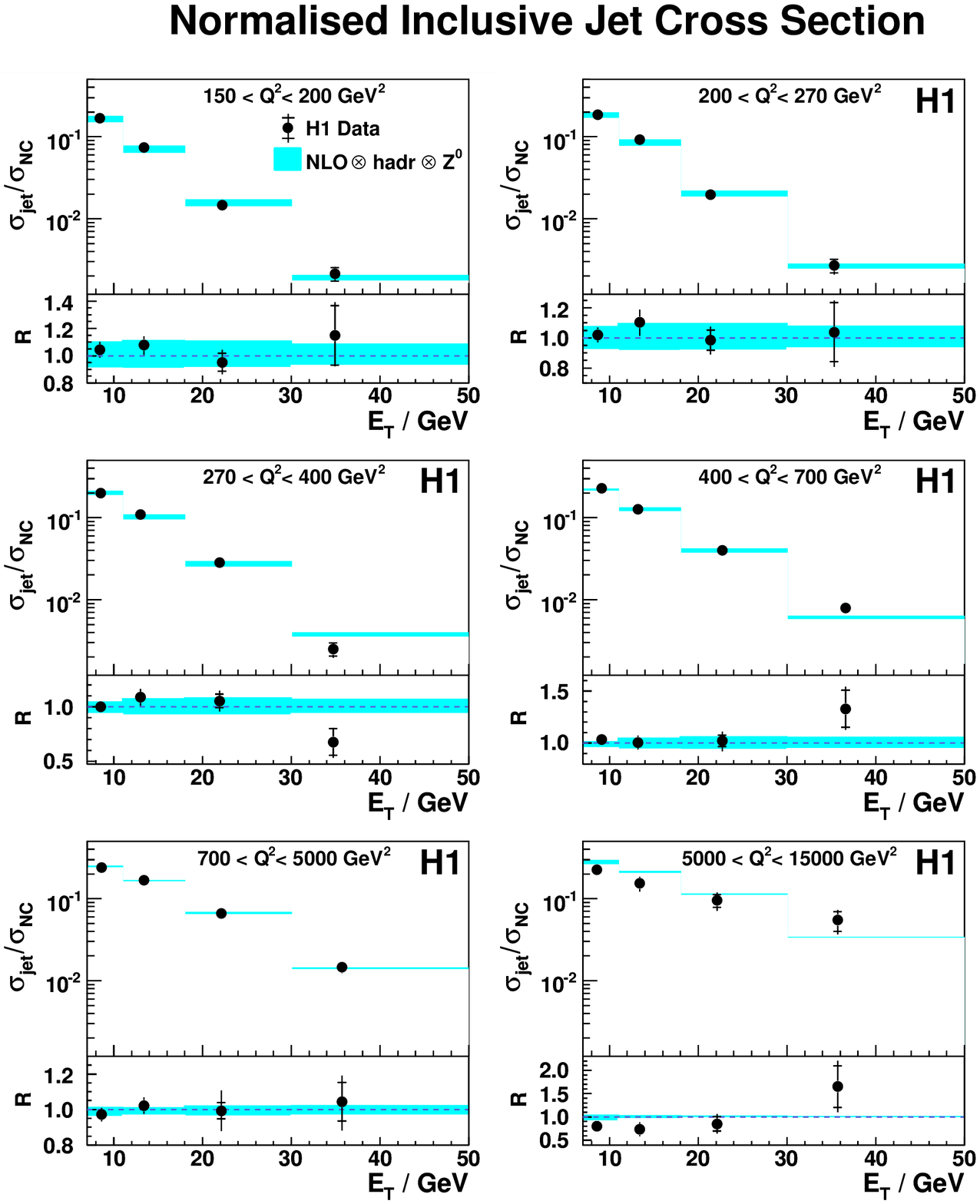}
\caption{The normalised inclusive jet cross section
 as a function of $E_T$ for six regions of $Q^2$.
The data, presented with statistical errors (inner bars)
 and total errors (outer bars),
 are compared with the results of the NLOJET++ and DISENT programs, corrected
for hadronisation effects and $Z^0$ boson exchange. 
The bands show the theoretical
uncertainty associated with the renormalisation and factorisation scales and the hadronisation correction.
In addition to the differential cross section, the ratio $R=\sigma_{\rm data}/\sigma_{\rm theory}$ is shown.
The band around $R=1$ displays the relative error of the theory calculation.
}
\label{fig:ratiodouble}
\end{figure}

The measured cross sections, corrected for detector
and radiative QED effects, are presented as single
and double differential distributions in figures \ref{fig:sigmasingle}-\ref{fig:ratiodouble}.
The data points are shown at the average value of the $Q^2$ or $E_T$ in each bin.
The results are compared to the perturbative QCD predictions in NLO with $\alpha_s(M_Z) = 0.118$,
 taking into account hadronisation effects and $Z^0$ boson exchange as explained in section~\ref{sec:nlo}.

The single differential inclusive jet cross sections, defined for events with inelasticity $0.2<y<0.7$
 and jets with pseudorapidity $-1.0 < \eta^\mathrm{Lab} < 2.5$,
are shown in figure~\ref{fig:sigmasingle} as functions of $Q^2$ and $E_T$.
A good description of the data by the theory calculation is observed.

The double differential inclusive jet cross section is shown in figure~\ref{fig:sigmadouble}  as a function of $E_T$ in
six $Q^2$ bins in the range $150 < Q^2 < 15000\,\mathrm{ GeV^2} $.
The data are well described by the theory over the full $E_T$ and $Q^2$ ranges, with $\chi^2/\mathrm{ndf} = 16.7/24$,
 taking only experimental errors into account.
The agreement is also good when $Q$ instead of $E_T$ is used in the calculation as renormalisation scale ($\chi^2/\mathrm{ndf} = 24.0/24$).

For NC DIS events in the range $0.2<y<0.7$ and in a given $Q^2$ bin
the normalised inclusive jet cross section is defined as the average number of jets within $-1.0 < \eta^\mathrm{Lab} < 2.5$
per event.
Figure~\ref{fig:ratiodouble} shows the normalised inclusive jet cross section as a function of $E_T$ in six $Q^2$ bins.
The NLO calculation gives a good description of the data in the full $E_T$ and $Q^2$ range.   
Compared with the inclusive jet cross section, the normalised 
inclusive jet cross section  exhibits a smaller experimental uncertainty.

\subsection{Extraction of the strong coupling}
The QCD predictions for jet production depend on $\alps$ and on 
the gluon and the quark density functions of the proton.
Using the present jet cross section measurements and the parton density functions from
 global analyses of inclusive deep-inelastic scattering and other data, $\alps$ is determined.

QCD predictions of the jet cross sections are calculated as a function of $\alpha_s(\mu_r=E_T)$
with the fastNLO package.
The cross sections are determined using the CTEQ6.5M proton PDFs
and hadronisation correction factors as described in section~\ref{sec:nlo}.
Measurements and theory predictions are used to calculate a $\chi^2(\alps)$ with the Hessian method, where
parameters representing systematic shifts of detector related observables are left free in the fit.
The experimental shifts (model dependence of the correction factors, positron energy scale, positron azimuth,
  hadronic final state energy scale and luminosity) found by the fit are consistent with the quoted uncertainties.
This method fully takes into 
account correlations of experimental uncertainties~\cite{chifit} and
has also been used in global data analyses~\cite{Botje:1999dj,Barone:1999yv} and
in previous H1 publications~\cite{Adloff:2000tq,Adloff:2000dp},
where a detailed description can be found.
The experimental uncertainty of $\alps$ is defined by 
that change in $\alps$ which gives an increase in $\chi^2$ of one unit with respect to the minimal value.
The theory error is estimated by adding in quadrature the deviation of $\alps$ from the central value when the fit
 is repeated with independent variations of the renormalisation scale, 
 the factorisation scale and the hadronisation correction factor. 

\begin{figure}
\center
\includegraphics[width=1.0\textwidth]{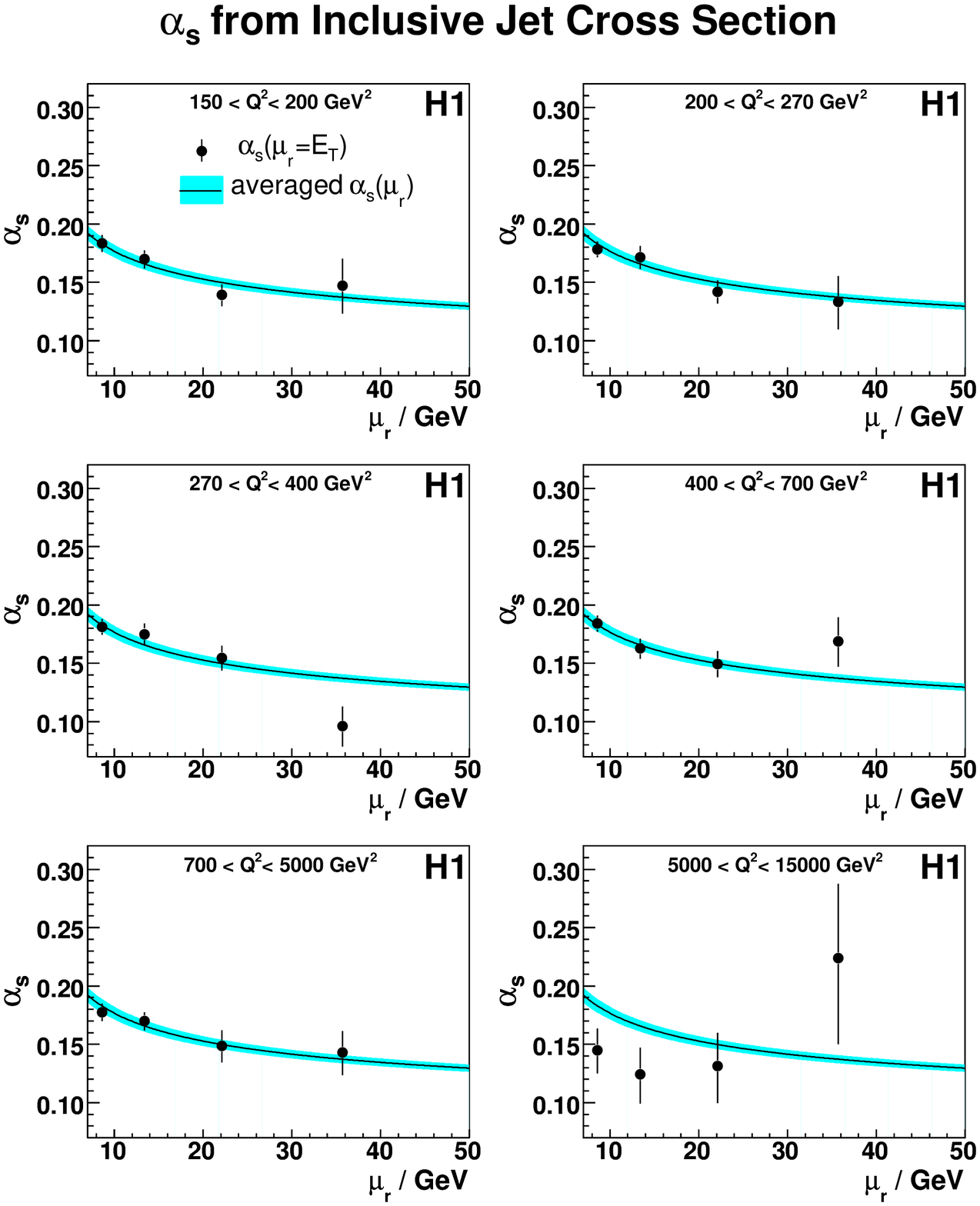}
\caption{
Results for the fitted values of $\alpha_s(E_T)$ using the inclusive jet cross section for six regions of $Q^2$.
The error bar denotes the uncorrelated experimental uncertainty for each fitted value.
The solid line shows the two loop solution of the renormalisation group equation
evolving the averaged $\alpha_s(M_Z)$ from all determinations, with the band 
 denoting the correlated experimental uncertainty.
 }
\label{fig:alphasmany}
\end{figure}

First, individual fits of $\alpha_s$ to each of the 24 measurements of the double differential inclusive jet cross sections
(presented in figure~\ref{fig:sigmadouble}) are made.
The resulting $\alpha_s(E_T)$ are shown in figure~\ref{fig:alphasmany}, for all bins.
These determinations demonstrate the property of asymptotic freedom of QCD and are in agreement
with the predicted scale dependence of $\alpha_s$.
The $\alpha_s$ values at the scale $E_T$ can also be related to the value of the strong coupling at the $Z^0$ mass 
$\alpha_s(M_Z)$ using the renormalisation group equation at two loops.
All 24 measurements are then used in a common fit of the strong coupling, 
which yields
\begin{eqnarray}
\alpha_s(M_Z) = 0.1179 ~\pm 0.0024\,\mathrm{(exp.)}
~ ^{+0.0052}_{-0.0032}\,\mathrm{(th.)}~ \pm 0.0028\,\mathrm{(pdf)}~ ,
\label{alphas1}
\end{eqnarray}
\noindent
with a fit quality: $\chi^2/\mathrm{ndf} = 20.2/23$.
The dominating experimental uncertainty is due to the LAr energy scale and the model dependence of the detector corrections.
 The renormalisation scale variation is the main contribution to the theory uncertainty, 
which dominates the overall uncertainty of this $\alpha_s$ determination.
The fit is repeated with $Q$ instead of $E_T$ as an alternative choice of renormalisation scale.
It yields a larger but compatible value of the strong coupling
 $\alpha_s(M_Z) = 0.1230 \pm 0.0028\,\mathrm{(exp.)}~ ^{+0.0036}_{-0.0054}\,\mathrm{(scale)} $
with $\chi^2/\mathrm{ndf} = 25.2/23$.
The quoted scale error corresponds to the variation of the renormalisation scale as described in section~\ref{sec:nlo}.

The global fit of the CTEQ6.5M PDFs was made assuming $\alpha_s(M_Z)=0.118$.
In order to test whether this value of $\alpha_s(M_Z)$ biases the results obtained using the nominal method presented above,
a method, similar to the one used in~\cite{Chekanov:2002be}, is employed using
the PDFs from the CTEQ6AB series, which were
obtained from global fits assuming different values for $\alpha_s(M_Z)$.
The cross section as a function of the strong coupling is interpolated with a polynomial and this
 interpolation is used to determine the best fit of the strong coupling to the data.
The result obtained with this alternative fit method is found to be compatible within $0.3$ standard deviations
 of the experimental error with the value from the nominal method.
Hence there is no indication for a bias due to the value of the strong coupling assumed for the CTEQ6.5M PDFs. 

\begin{figure}
\center
\includegraphics[width=1.0\textwidth]{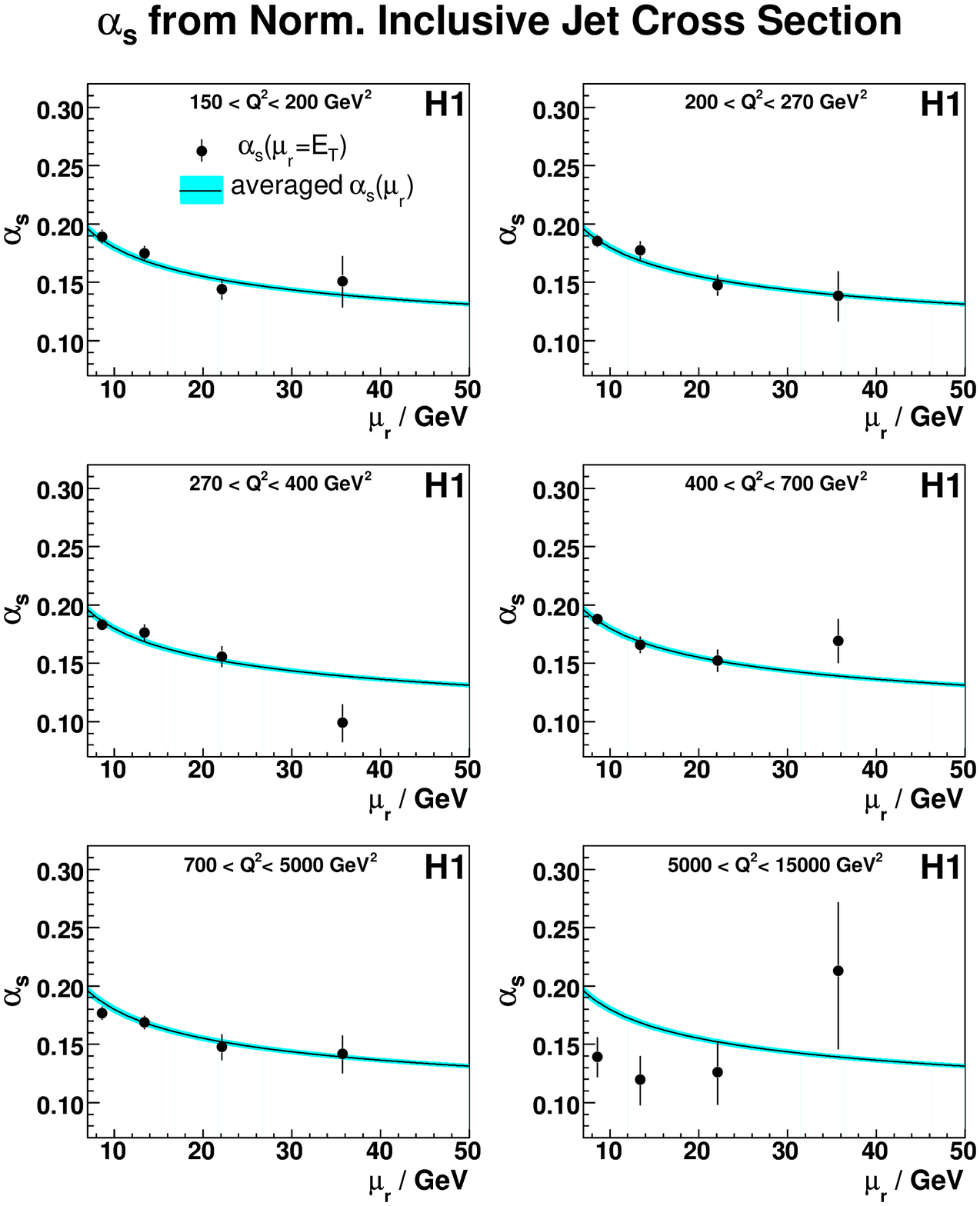}
\caption{
Results for the fitted values of $\alpha_s(E_T)$ for six regions of $Q^2$ using the normalised inclusive jet cross section.
The error bar denotes the uncorrelated experimental uncertainty for each fitted value.
The solid line shows the two loop solution of the renormalisation group equation
evolving the averaged $\alpha_s(M_Z)$ from all determinations, with the band 
 denoting the correlated experimental uncertainty.
 }
\label{fig:alphasmanynorm}
\end{figure}

\begin{figure}
\center
\includegraphics[width=1.0\textwidth]{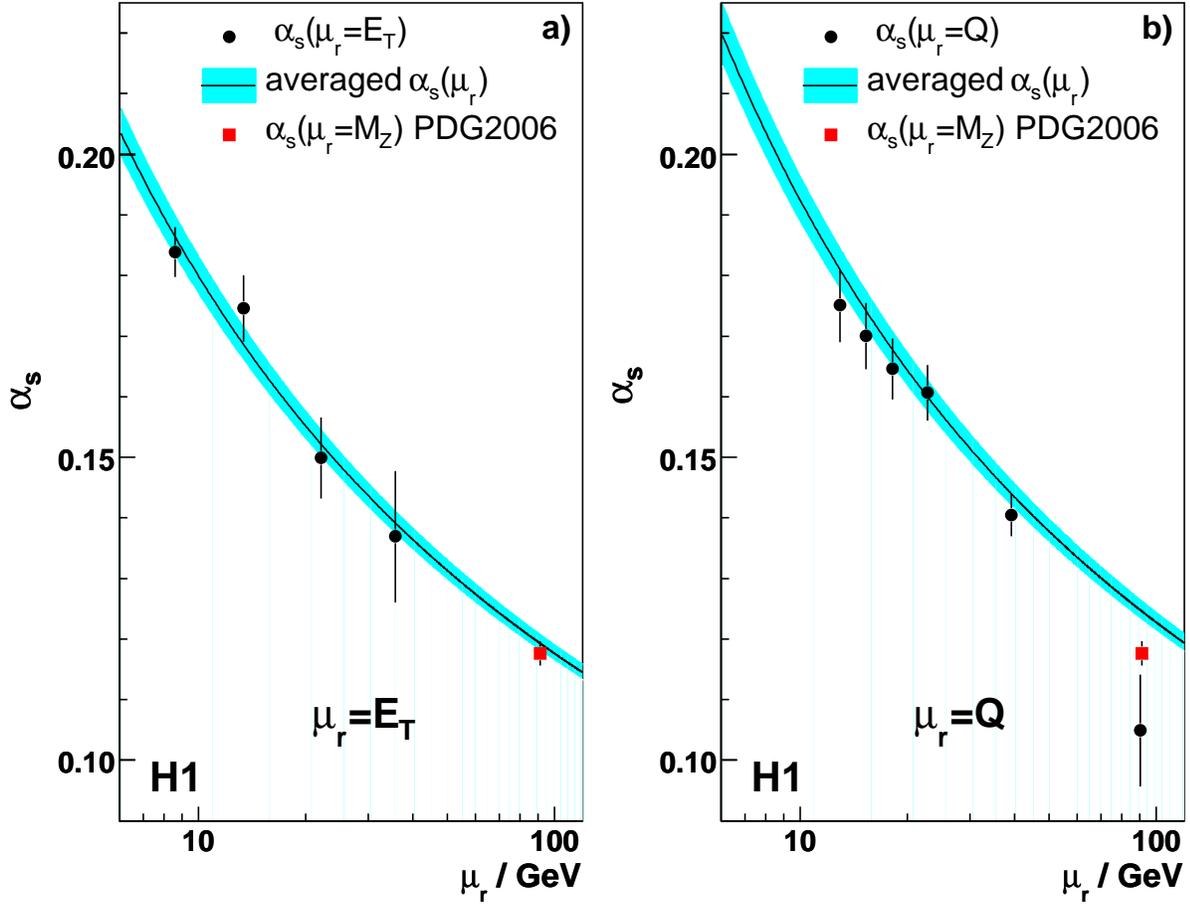}
\caption{
Results for the fitted values of a) $\alpha_s(\mu_r=E_T)$ averaged over all $Q^2$ regions, and b) 
 $\alpha_s(\mu_r=Q)$ averaged over all $E_T$ regions.
The error bars denote the total experimental uncertainty for each data point.
The solid curve shows the result of evolving $\alpha_s(M_Z)$ averaged from all $Q^2$ and $E_T$ regions, 
 with the band denoting the total experimental uncertainty.
The world average from PDG is also shown.
 }
\label{fig:theoet}
\end{figure}

The measurements of the normalised inclusive jet cross section are also used to extract the strong coupling using
 the nominal fit method.
The resulting $\alpha_s(E_T)$ are shown 
in figure~\ref{fig:alphasmanynorm}, for all bins.
As the results are consistent over the whole
range of $Q^2$ and $E_T$, combined fits are made to 
groups of data points.
To study the scale dependence of $\alpha_s$, the six data points with different $Q^2$ at a given
$E_T$ are used together, and four values of $\alpset$ are extracted.
The results are shown in figure~\ref{fig:theoet}a,
 where the running of the strong coupling  is also clearly observed.
Finally, all 24 measurements are used in a common fit of the strong coupling, 
which yields
\begin{eqnarray}
\alpha_s(M_Z) = 0.1193 ~\pm 0.0014\,\mathrm{(exp.)}
~ ^{+0.0047}_{-0.0030}\,\mathrm{(th.)}~ \pm 0.0016\,\mathrm{(pdf)}~ ,
\end{eqnarray}
with a fit quality of $\chi^2/\mathrm{ndf} = 28.7/23$.
This result is compatible within errors with the value from the inclusive jet cross sections quoted in \ref{alphas1}.
The normalisation gives rise to cancellations of systematic effects, which lead to improved experimental and PDF uncertainties.
This determination of $\alpha_s(M_Z)$ is consistent with the world 
average $\alpha_s(M_Z)=0.1176 \pm 0.0020$ ~\cite{Yao:2006px} and with the previous H1
determination from inclusive jet production measurements~\cite{Adloff:2000tq}.
In figure~\ref{fig:theoet}b the running of the strong coupling is studied
 using the alternative scale $Q$ instead of $E_T$:
the four data points at a given $Q^2$ are used together, and six values of $\alpha_s(Q)$ are extracted.

The dominating theory error can be reduced at the expense of a larger experimental uncertainty
by restricting the data points included in the fit to those at higher values of $Q^2$.
The smallest total uncertainty is obtained by a combined fit
 of the normalised inclusive jet cross section for $700<Q^2<5000\gev^2$,
\begin{eqnarray}
\alpha_s(M_Z) = 0.1171 ~\pm 0.0023\,\mathrm{(exp.)}
~ ^{+0.0032}_{-0.0010}\,\mathrm{(th.)}~ \pm 0.0010\,\mathrm{(pdf)}~,
\end{eqnarray}
\noindent
with a fit quality of $\chi^2/\mathrm{ndf} = 1.2/3$.

\section{Conclusion}

Measurements of inclusive jet cross sections 
in the Breit frame in deep-inelastic positron-proton scattering in the range $150<Q^2<15000\gev^2$ are presented,
together with the normalised inclusive jet cross sections, defined as the ratio of 
the inclusive jet cross section to the NC~DIS cross section within the given phase space.
Calculations at NLO QCD, corrected for hadronisation
effects, provide a good description of
the single and double differential cross sections as functions of the jet transverse energy $E_T$ and $Q^2$.
The strong coupling $\alpha_s$ is determined from a fit of the NLO prediction to the measurements.
The experimentally most precise determination of $\alpha_s(M_Z)$ is derived from
the measurement of the normalised inclusive jet cross section:
\begin{eqnarray}
\nonumber
\alpha_s(M_Z) = 0.1193 ~\pm 0.0014\,\mathrm{(exp.)}
~ ^{+0.0047}_{-0.0030}\,\mathrm{(th.)}~ \pm 0.0016\,\mathrm{(pdf)}~ .
\end{eqnarray}
\noindent
Additionally, the PDF uncertainty is significantly reduced compared to the determination from the inclusive jet cross section.
The dominating source of error is the renormalisation scale
 dependence which is used to estimate the effect of missing higher orders beyond NLO in the pQCD prediction.
This result shows a level of experimental precision competitive with $\alpha_s$ determinations from 
other recent jet production measurements at HERA~\cite{Chekanov:2006yc} and those from $e^+ e^-$ data~\cite{Abbiendi:2005vd} 
and is in good agreement with the world average.


\section*{Acknowledgements}

We are grateful to the HERA machine group whose outstanding
efforts have made this experiment possible.
We thank
the engineers and technicians for their work in constructing and
maintaining the H1 detector, our funding agencies for
financial support, the
DESY technical staff for continual assistance
and the DESY directorate for support and for the
hospitality which they extend to the non DESY
members of the collaboration.


\clearpage
\begin{center}
\scriptsize \sf
\setlength{\tabcolsep}{1pt}
\begin{tabular}{crcccr}
\hline
\\[-3mm]
bin number & \multicolumn{5}{c}{ corresponding $Q^2$ range} \\
\hline
\\[-3mm]
\,\, 1 \,\, & $150$&$<$&$Q^2$&$<$&$200\gev^2$ \\
\,\, 2 \,\, & $200$&$<$&$Q^2$&$<$&$270\gev^2$ \\
\,\, 3 \,\, & $270$&$<$&$Q^2$&$<$&$400\gev^2$ \\
\,\, 4 \,\, & $400$&$<$&$Q^2$&$<$&$700\gev^2$ \\
\,\, 5 \,\, & $700$&$<$&$Q^2$&$<$&$5000\gev^2$ \\
\,\, 6 \,\, &$5000$&$<$&$Q^2$&$<$&$15000\gev^2$ \\
\\[-3mm]
\hline
\end{tabular}
\hskip20mm
\begin{tabular}{cl}
\hline
\\[-3mm]
bin letter &  corresponding $E_T$ range \\
\hline
\\[-3mm]
\,\, a \,\, &  $ \;\,7 < E_T < 11\gev$ \\ 
\,\, b \,\, &  $ 11 < E_T < 18\gev$ \\ 
\,\, c \,\, &  $ 18 < E_T < 30\gev$ \\ 
\,\, d \,\, &  $ 30 < E_T < 50\gev$ \\ 
\\[-3mm]
\hline
\end{tabular}
\end{center}

\begin{table}[h!]
\tiny \sf
\centering
\begin{tabular}{rrr r r r r r r r r} 
\hline
\multicolumn{11}{c}{ } \\
\multicolumn{11}{c}{\normalsize Inclusive jet cross section in bins of $Q^2$ and $E_T$} \\
\multicolumn{11}{c}{ } \\
\hline
 & & &  & total & total &
\multicolumn{4}{c}{\underline{\hspace*{0.9cm}single contributions to correlated uncertainty\hspace*{0.9cm}}}
&\\
bin & cross & statistical & total & uncorrelated & correlated  &
model dep. & positron  & positron & HFS hadr. & hadronis. \\
& section & uncert. & uncertainty  & uncertainty & uncertainty
& detector corr. & energy scale & polar angle & energy scale &  correct. \\
& (in pb) & (in percent) & (in percent) & (in percent) & (in percent)
& (in percent) & (in percent) & (in percent) & (in percent) & factor \\
\hline
 1 a&   73.81 &  2.1 &  6.8 &  4.9 &  4.7 &  3.5 &  0.7 &  0.4 &  2.6 & 1.076 \\
 1 b&   32.44 &  3.1 &  7.7 &  5.8 &  5.1 &  2.7 &  0.8 &  0.3 &  3.9 & 1.035 \\
 1 c&    6.40 &  7.0 & 10.4 &  8.8 &  5.6 &  1.7 &  0.3 &  0.6 &  5.1 & 1.032 \\
 1 d&    0.94 & 18.9 & 21.0 & 19.9 &  6.5 &  0.8 &  1.1 &  0.9 &  6.1 & 1.065 \\
\hline
 2 a&   58.06 &  2.2 &  6.3 &  4.6 &  4.3 &  3.1 &  0.1 &  0.4 &  2.6 & 1.075 \\
 2 b&   28.85 &  3.1 &  9.5 &  7.0 &  6.5 &  4.9 &  0.3 &  0.1 &  3.9 & 1.034 \\
 2 c&    6.16 &  6.8 & 10.5 &  8.8 &  5.7 &  1.0 &  0.7 &  0.0 &  5.4 & 1.040 \\
 2 d&    0.85 & 18.9 & 21.7 & 20.3 &  7.5 &  2.2 &  1.2 &  0.8 &  6.9 & 1.044 \\
\hline
 3 a&   55.16 &  2.2 &  5.6 &  4.1 &  3.8 &  2.6 &  0.2 &  0.1 &  2.4 & 1.085 \\
 3 b&   30.45 &  2.9 &  8.7 &  6.4 &  5.9 &  4.2 &  0.1 &  0.1 &  3.8 & 1.032 \\
 3 c&    7.87 &  6.0 & 10.7 &  8.6 &  6.3 &  3.1 &  0.1 &  0.1 &  5.3 & 1.029 \\
 3 d&    0.69 & 18.1 & 21.5 & 19.9 &  8.2 &  2.0 &  0.7 &  1.1 &  7.7 & 1.039 \\
\hline
 4 a&   48.50 &  2.3 &  5.0 &  3.8 &  3.3 &  1.8 &  0.2 &  0.5 &  2.4 & 1.093 \\
 4 b&   26.81 &  3.0 &  8.0 &  5.9 &  5.3 &  3.8 &  0.2 &  0.4 &  3.5 & 1.035 \\
 4 c&    8.46 &  5.4 & 10.7 &  8.4 &  6.6 &  3.1 &  0.4 &  0.1 &  5.6 & 1.025 \\
 4 d&    1.69 & 13.3 & 16.4 & 14.9 &  6.9 &  3.2 &  0.5 &  0.5 &  5.9 & 1.035 \\
\hline
 5 a&   43.02 &  2.4 &  5.2 &  3.9 &  3.5 &  1.7 &  0.2 &  0.8 &  2.5 & 1.103 \\
 5 b&   30.23 &  2.9 &  6.2 &  4.7 &  4.0 &  2.5 &  0.4 &  0.7 &  2.7 & 1.040 \\
 5 c&   11.88 &  4.5 & 13.6 & 10.1 &  9.1 &  7.9 &  0.2 &  0.6 &  4.3 & 1.038 \\
 5 d&    2.63 & 10.3 & 16.5 & 13.7 &  9.2 &  6.3 &  0.3 &  0.9 &  6.5 & 1.046 \\
\hline
 6 a&    1.79 & 10.8 & 12.8 & 11.8 &  5.0 &  0.8 &  4.0 &  0.3 &  1.8 & 1.083 \\
 6 b&    1.23 & 13.4 & 22.7 & 18.5 & 13.1 & 10.8 &  6.4 &  1.9 &  2.2 & 1.050 \\
 6 c&    0.76 & 17.8 & 27.2 & 22.9 & 14.7 & 10.8 &  9.2 &  2.0 &  2.7 & 1.029 \\
 6 d&    0.44 & 26.8 & 34.3 & 30.8 & 15.2 & 14.5 &  3.0 &  1.0 &  3.3 & 1.029 \\
\hline
\\
\\
\\
\hline
\\
\multicolumn{11}{c}{\normalsize Inclusive jet cross section in bins of $E_T$}\\
\\
\hline
   a&  281.43 &  0.9 &  5.7 &  4.0 &  4.1 &  2.8 &  0.2 &  0.3 &  2.5 & 1.084 \\
   b&  150.22 &  1.1 &  7.7 &  5.5 &  5.4 &  3.8 &  0.2 &  0.2 &  3.6 & 1.035 \\
   c&   41.70 &  2.1 &  9.6 &  7.0 &  6.6 &  4.1 &  0.1 &  0.2 &  5.1 & 1.033 \\
   d&    7.29 &  5.1 & 12.1 &  9.4 &  7.6 &  3.8 &  0.5 &  0.3 &  6.5 & 1.045 \\
\hline
\\
\\
\\
\hline
\\
\multicolumn{11}{c}{\normalsize Inclusive jet cross section in bins of $Q^2$}\\
\\
\hline
   1&  113.73 &  1.8 &  6.9 &  4.9 &  4.8 &  3.2 &  0.6 &  0.4 &  3.2 & 1.062 \\
   2&   94.06 &  1.9 &  7.1 &  5.1 &  5.0 &  3.5 &  0.2 &  0.3 &  3.2 & 1.060 \\
   3&   94.30 &  1.8 &  6.7 &  4.8 &  4.7 &  3.1 &  0.1 &  0.0 &  3.2 & 1.063 \\
   4&   85.41 &  1.9 &  6.2 &  4.5 &  4.3 &  2.6 &  0.1 &  0.4 &  3.1 & 1.067 \\
   5&   87.81 &  1.9 &  6.5 &  4.7 &  4.5 &  3.0 &  0.2 &  0.7 &  2.9 & 1.071 \\
   6&    4.24 &  8.2 & 15.1 & 12.1 &  9.1 &  7.0 &  5.0 &  1.1 &  2.2 & 1.061 \\
\hline
\end{tabular}
\rm \normalsize
\caption{
Results of the inclusive jet cross section measurement
using the inclusive $k_T$ algorithm in the Breit frame 
for the phase space $0.2<y<0.7$,  $7<E_T<50\gev$ and $-1.0 < \eta^\mathrm{Lab} < 2.5$.
The multiplicative hadronisation correction factor as applied to the NLO calculation is shown in the last column.
The contribution of $\pm 1.5\%$ from the luminosity measurement uncertainty is included in the total correlated uncertainty.
\label{tab:res_incl}}
\end{table}

\begin{table}[htp]
\tiny \sf
\centering
\begin{tabular}{rrr r r r r r r r r} 
\hline
\multicolumn{11}{c}{ } \\
\multicolumn{11}{c}{\normalsize Normalised inclusive jet cross section in bins of $Q^2$ and $E_T$}\\
\multicolumn{11}{c}{ } \\
\hline
 & & &  & total & total &
\multicolumn{4}{c}{\underline{\hspace*{0.9cm}single contributions to correlated uncertainty\hspace*{0.9cm}}}
&\\
bin & normalised & statistical & total & uncorrelated & correlated  &
model dep. & positron  & positron & HFS hadr. & hadronis. \\
 & cross & uncert. & uncertainty  & uncertainty & uncertainty
& detector corr. & energy scale & polar angle & energy scale &  correct. \\
& section & (in percent) & (in percent) & (in percent) & (in percent)
& (in percent) & (in percent) & (in percent) & (in percent) & factor \\
\hline
 1 a&   0.168 &  2.1 &  5.4 &  4.0 &  3.7 &  2.8 &  0.6 &  0.3 &  1.8 & 1.076 \\
 1 b&   0.074 &  3.1 &  6.4 &  4.9 &  4.1 &  2.0 &  0.8 &  0.6 &  3.1 & 1.035 \\
 1 c&   0.015 &  7.0 &  9.5 &  8.3 &  4.7 &  1.0 &  0.4 &  0.4 &  4.3 & 1.032 \\
 1 d&   0.002 & 18.9 & 20.5 & 19.7 &  5.6 &  0.0 &  1.1 &  0.3 &  5.3 & 1.065 \\
\hline
 2 a&   0.184 &  2.2 &  4.7 &  3.5 &  3.2 &  2.2 &  0.6 &  0.2 &  1.6 & 1.075 \\
 2 b&   0.092 &  3.1 &  7.9 &  5.9 &  5.3 &  4.0 &  0.8 &  0.4 &  2.9 & 1.034 \\
 2 c&   0.020 &  6.8 &  9.5 &  8.2 &  4.8 &  0.1 &  1.2 &  0.4 &  4.4 & 1.041 \\
 2 d&   0.003 & 18.9 & 21.3 & 20.1 &  7.0 &  3.1 &  0.8 &  1.1 &  5.9 & 1.044 \\
\hline
 3 a&   0.199 &  2.2 &  4.2 &  3.2 &  2.8 &  1.9 &  0.5 &  0.1 &  1.3 & 1.085 \\
 3 b&   0.110 &  2.9 &  7.1 &  5.3 &  4.7 &  3.5 &  0.5 &  0.0 &  2.7 & 1.032 \\
 3 c&   0.028 &  6.0 &  9.2 &  7.7 &  5.1 &  2.4 &  0.3 &  0.2 &  4.2 & 1.030 \\
 3 d&   0.003 & 18.1 & 20.7 & 19.4 &  7.0 &  1.3 &  0.9 &  1.2 &  6.5 & 1.039 \\
\hline
 4 a&   0.228 &  2.3 &  3.8 &  3.0 &  2.4 &  1.3 &  0.5 &  0.1 &  1.2 & 1.093 \\
 4 b&   0.126 &  3.0 &  6.7 &  5.1 &  4.3 &  3.3 &  0.6 &  0.3 &  2.3 & 1.035 \\
 4 c&   0.040 &  5.4 &  9.3 &  7.5 &  5.5 &  2.7 &  0.8 &  0.7 &  4.4 & 1.025 \\
 4 d&   0.008 & 13.3 & 15.5 & 14.4 &  5.7 &  2.7 &  0.1 &  0.3 &  4.8 & 1.035 \\
\hline
 5 a&   0.239 &  2.4 &  3.7 &  3.0 &  2.3 &  0.3 &  0.7 &  0.2 &  1.5 & 1.103 \\
 5 b&   0.168 &  2.9 &  4.5 &  3.6 &  2.6 &  1.1 &  0.7 &  0.2 &  1.6 & 1.040 \\
 5 c&   0.066 &  4.5 & 11.4 &  8.6 &  7.5 &  6.5 &  0.6 &  0.1 &  3.3 & 1.038 \\
 5 d&   0.015 & 10.3 & 14.8 & 12.7 &  7.5 &  4.9 &  0.4 &  0.3 &  5.5 & 1.046 \\
\hline
 6 a&   0.225 & 10.8 & 11.7 & 11.2 &  3.2 &  0.4 &  1.8 &  0.6 &  1.7 & 1.083 \\
 6 b&   0.154 & 13.4 & 20.6 & 17.3 & 11.2 &  9.5 &  4.7 &  1.9 &  1.8 & 1.050 \\
 6 c&   0.095 & 17.8 & 25.2 & 21.7 & 12.7 &  9.6 &  7.6 &  2.2 &  2.2 & 1.029 \\
 6 d&   0.055 & 26.8 & 33.5 & 30.3 & 14.2 & 13.2 &  4.0 &  0.8 &  2.9 & 1.029 \\
\end{tabular}
\rm \normalsize
\caption{Results of the normalised inclusive jet cross section measurement, i.e.\ the average number of jets with $7\gev<E_T<50\gev$ and $-1.0 < \eta^\mathrm{Lab} < 2.5$  per NC DIS event for the phase space $0.2<y<0.7$.
The multiplicative hadronisation correction factor as applied to the NLO calculation is shown in the last column.
The $Q^2$, $E_T$ binning scheme is displayed in Table~\ref{tab:res_incl}.
\label{tab:res_ratio}}

\end{table}

\clearpage

\end{document}